# Combinatorial complexity and dynamical restriction of network flows in signal transduction

James R. Faeder*, Michael L. Blinov&, Byron Goldstein & William S. Hlavacek

*Theoretical Biology and Biophysics Group, Theoretical Division, Los Alamos National Laboratory, Mail Stop K710, Los Alamos, New Mexico 87545, USA*

* To whom correspondence should be addressed.  E-mail: faeder@lanl.gov

**Abstract:** The activities and interactions of proteins that govern the cellular response to a signal generate a multitude of protein phosphorylation states and heterogeneous protein complexes.  Here, using a computational model that accounts for 307 molecular species implied by specified interactions of four proteins involved in signalling by the immunoreceptor FcεRI, we determine the relative importance of molecular species that can be generated during signalling, chemical transitions among these species, and reaction paths that lead to activation of the protein tyrosine kinase (PTK) Syk.  By all of these measures and over 2- and 10-fold ranges of model parameters — rate constants and initial concentrations — only a small portion of the biochemical network is active.  The spectrum of active complexes, however, can be shifted dramatically, even by a change in the concentration of a single protein, which suggests that the network can produce qualitatively different responses under different cellular conditions and in response to different inputs.  Reduced models that reproduce predictions of the full model for a particular set of parameters lose their predictive capacity when parameters are varied over 2-fold ranges.

## 1. Introduction

Cell signalling, the biochemical process through which cells sense and respond to their environment, involves an array of proteins, which include receptors, kinases, and adaptors, components of proteins, such as sites of phosphorylation, and other biomolecules [1]. Early signalling events triggered by receptors in eukaryotic cells usually involve the formation of heterogeneous protein complexes in the vicinity of the cell membrane [2-4]. This process of complex formation is complicated because a typical signalling protein contains multiple sites that may be modified (e.g., phosphorylated) and that have the potential to bind other proteins or lipids. In addition, the modification or binding state of a protein can regulate its binding and enzymatic activities. Thus, signalling can generate a combinatorially large number protein states and complexes with different potentials to generate further signals [4-8]. For example, a protein that contains 10 amino acid residues subject to the activities of kinases and phosphatases theoretically has $2^{10}$=1024 states of phosphorylation. If the protein forms homodimers, the number of chemically distinct phosphorylation states is 524,800, a number that might exceed the total amount of this protein in the cell. This combinatorial complexity has been largely ignored by both experimentalists and modellers and is a major barrier to predictive understanding of signal transduction.

Experimental resolution of protein states and complexes is usually limited to a small number of sites and interactions, but rapidly advancing proteomic technologies are likely to provide a wealth of more detailed information about signalling complexes in the near future [9-13]. A number of studies already confirm that a diverse range of molecular complexes arise during signal transduction [14-17]. Because the full spectrum of protein states and complexes is difficult to enumerate, let alone understand, computational modelling will play an important role in interpreting such data and assessing the functional significance of specific interactions and complexes [8]. Key questions to be addressed include whether networks favour the formation of specific complexes from the multitude of potential complexes, and, if so, how these favoured complexes affect signalling outcomes.

Few biochemical network models of signalling developed so far encompass the breadth of states and complexes (which we refer to collectively as molecular species) required to address these questions. Instead, most models, given a particular set of proteins and interactions, make additional (usually implicit) assumptions that exclude the vast majority of possible species from consideration. An example is the model of epidermal growth factor receptor signalling that was developed by Kholodenko et al. [18] and extended by several other groups (for example, see Refs. [19] and [20]). The original model includes six proteins and tracks 25 species, but lifting implicit assumptions in the model raises the number to hundreds or thousands of species, depending on mechanistic assumptions, even without the introduction of new rate constants or other parameters [8]. While such models have provided valuable insights into signalling mechanisms, they are not suitable for addressing the questions of whether and how signalling networks favour specific complexes, which requires models that consider the full spectrum of possible species.

Here, we analyze the specificity of complex formation in a network model for early events in signalling by the high-affinity receptor for IgE antibody (FcεRI), a key initiator of allergic reactions [21]. The model has been shown to make accurate predictions of a number of experimental observations [22, 23]. Here, we characterize the distribution of network activity in terms of individual species, reactions, and reaction sequences or paths. We then examine how the spread of network activity is affected when model parameters are randomly varied, which corresponds to changing the initial state of the cell that is receiving the signal. We also explore the possibility of developing an accurate reduced model by removing nonessential species from the reaction network. The results indicate that while only a small fraction of complexes, reactions, and paths is active for a particular cellular state, which elements are active depends strongly on the initial state of the cell. Thus, to capture the full range of signalling behaviours, a model must account for many more molecular complexes than just those that are favoured in any particular cellular state.

## 2. Methods

*Network Model.* The network model analyzed in this study was developed in earlier work [23]. It includes just four components (Fig. 1A)—the FcεRI receptor, a bivalent ligand that binds to a single site on FcεRI, and the protein tyrosine kinases Lyn and Syk—but, in a vivid illustration of combinatorial complexity, encompasses 354 species coupled through a biochemical network of 3680 unidirectional reactions. The size of the network is probably typical of cascades involving a comparable number of proteins [8]. Despite its large size, this network can be characterized by a relatively small set of parameters—the initial concentrations of the four components and 21 chemical rate constants [23]. A series of events (Fig. 1B) couples binding of the ligand to activation of Syk [24-26], which is required for downstream signalling events and cellular responses, such as calcium mobilization and release of histamine from mast cells [27, 28]. Figure 1C displays one of a multitude of possible sequences of individual reaction steps starting from an unmodified receptor and leading to a dimer of receptors containing fully-phosphorylated Syk. At each step along this path many alternative branches are possible, as indicated by the highlighted state in Fig. 1C and quantified by the distribution in the number of reactions a species containing a dimer of receptors can undergo (Fig. 1D).

*Time courses.* Elementary mass action kinetics give rise to a system of coupled ordinary differential equations (ODEs) that describe the time evolution of the species concentrations following the addition of ligand. 307 species, connected through a network of 2326 unidirectional reactions, are relevant when the ligand binds the receptor irreversibly, which is effectively the case for the ligand considered here, a chemically cross-linked dimer of IgE antibodies [24]. The rules used to generate the network along with the default values of the component concentrations and rate parameters that characterize the rat basophilic leukaemia cell line RBL-2H3 are given in Figure 1 and Table 1 of [23]. The BioNetGen software package was used to construct the model based on these rules and parameters and to perform calculations [29]. The model and the software are available at http://cellsignaling.lanl.gov.

*Measure of importance.* In an ordered distribution of concentration or flux over a set of network elements (e.g., a set of species or reactions), we label an element "important" if it is one of the top *n* elements that account for 95% of the distributed quantity. The choice of cutoff is arbitrary, but for an even distribution over the network elements, the fraction of important network elements should equal the cutoff value. When the fraction of important elements is smaller than the cutoff value, the distribution can be considered skewed.

*Syk activation paths.* We define an activation path as a sequence of reaction events by which a molecular component of the model is transformed from an inactive state into an active one. Here, we analyze the paths that transform an unphosphorylated Syk molecule in the cytosol into an autophosphorylated Syk molecule associated with a receptor dimer complex (Syk*). As described in more detail in [Appendix 1](), we use a deterministic algorithm to enumerate paths as a function of the path length and a stochastic algorithm to compute the relative contribution of each path to the rate of Syk* production.

*Parameter set ensembles.* To determine the possible effect of the initial cellular state on the distribution of network activity, we generated two ensembles of 5000 randomly scaled sets of parameter values, referred to as the 2x and 10x ensembles.
Each new parameter set is produced by scaling each of the parameters in the original model (except the two rate constants $k_{+1}$ and $k_{+2}$, which characterize ligand-receptor binding) by an amount $x^p$, where $p$ is a uniformly distributed random variable on the interval [-1,1] chosen separately for each parameter. The ensembles are labelled by their $x$ value, $x=2$ or $x=10$. For each new parameter set generated, the time evolution of the 354 chemical species is obtained as described above. Varying the parameters has the effect of changing length of time required to achieve steady state. A fixed time of 100 s was chosen for computing the Syk* distribution and the reaction rate distribution in order to sample a range of times relative to the attainment of steady state. A later sampling time was chosen for the activation path distribution to ensure the accuracy of the path sampling method (see Appendix 1), the validity of which depends on steady-state conditions. The ligand-receptor binding constants $k_{+1}$ and $k_{+2}$ were fixed so as to focus our analysis on parameters that govern the intracellular signalling dynamics. The ligand concentration is 1 nM in the unscaled parameter set.

*Model Reduction.* We used an optimization procedure based on deleting species from the full network model to find the smallest network that will reproduce the time course of the full model for a set of observed quantities to within a specified error. When a species is deleted from the network, all reactions associated with that species are also removed, but none of the remaining reactions or reaction rate constants is changed. The objective function used to test the fitness of a reduced model is the root-mean squared (RMS) of the relative error computed over all quantities and time points. The six quantities, which correspond to observable properties that either have been or could be measured for this system, are FcεRIβ ITAM phosphorylation, FcεRIγ ITAM phosphorylation, Syk linker region phosphorylation, Syk kinase activation loop phosphorylation, association of Lyn with the unphosphorylated FcεRIβ subunit, and association of Lyn with the phosphorylated FcεRIβ ITAM measured at 1, 10, 100, and 1000 s after addition of ligand. Details of the optimization algorithm are presented in [Appendix 2](Appendix 2).

## 3. Results

As measures of the breadth of network activity we consider the distribution of activated Syk in receptor-containing complexes, the distribution of chemical reaction rates among reactions that share the same rate constant, and the distribution of fluxes among paths that lead to activated Syk. These distributions are obtained for a default set of parameters that characterize the rat basophilic leukaemia cell line (RBL-2H3) [23], for the default set with the Lyn concentration increased ten-fold, and finally for ensembles of parameter sets in which the default values are randomly varied over 2-fold and 10-fold ranges.

*Distribution of activated Syk (Syk\*).* A Syk molecule that is bound to a receptor and has been phosphorylated by a second Syk is considered to be activated. The 164 species that contain Syk\* represent chemically distinct output channels of the signalling model. We find that only a few of these channels dominate the distribution of Syk\* at all times following addition of ligand. The two most populated species, 354 and 207, contain more than 50% of the Syk\* (Fig. 2A), and 12 species contain more than 95% of the Syk\* (Fig. 2B, black bars). Although relatively few Syk\* species are populated, the composition of these species is heterogeneous (Fig. 2C), varying in the amount of associated Lyn and in the level of Lyn-mediated phosphorylation of Syk. For example, Species 354 contains two Lyn molecules and two Lyn-phosphorylated Syk\* molecules, whereas Species 207 contains no Lyn and neither of its two Syk\* molecules is Lyn phosphorylated. This heterogeneity may have functional consequences, because Lyn and Lyn-phosphorylated Syk contain binding sites for signalling molecules [30-34] including Cbl, the p85 subunit of phosphatidylinositol-3' kinase, and phospholipase C$\gamma$. As a result, molecules associated with Lyn-containing and Lyn-deficient Syk\* species can differ and the different signalling complexes have the potential to trigger distinct downstream signalling events.

The predicted distribution of Syk\* changes during the response to stimulation (Fig. 2A). The Lyn-containing complex, 354, exhibits faster initial kinetics than the Lyn-deficient complex, 207, but as receptor phosphorylation increases, the pool of free Lyn available to bind receptors is depleted [35], and 207 replaces 354 as the most abundant form of Syk*. Thus, the temporal redistribution of Syk* could have functional consequences if co-localization of Lyn and Syk has a strong effect on downstream signals.

    The predicted distribution of Syk* also depends on the initial state of a cell. As illustrated in Figure 2B, the distribution of Syk* can be shifted by a change in the concentration of a single component. Increasing the concentration of Lyn 10-fold causes a redistribution of Syk* into Lyn-containing complexes (Fig. 2B, red bars). The effect on Lyn-deficient states can be quite large: for example, the fraction of Syk* in Species 207 drops by more than a factor of 1000. Thus, a cellular response that depends on co-localization of Lyn and Syk could be upregulated (downregulated) by increasing (decreasing) the expression of Lyn. Unfortunately, without including additional components in the model, it is difficult to predict how co-localization would affect activity. For example, Lyn-containing Syk* complexes might upregulate Syk-dependent responses because Lyn binds the regulatory subunit of phosphatidylinositol-3' kinase (PI3K) [30], whose catalytic activity creates plasma membrane binding sites for a number of known Syk substrates [34]. On the other hand, Lyn-containing Syk* complexes might downregulate Syk-dependent responses because Lyn phosphorylation of Syk on Tyr-317 creates a binding site for the ubiquitin ligase Cbl, which marks Syk for degradation and may block the direct binding of PLC-$\gamma$ to other phosphotyrosine residues on Syk [33]

***Distribution of reaction rates.*** Another way to measure the importance of network elements is to examine rates of individual chemical reactions. The model is constructed by lumping together similar chemical transformations into classes described by a single rate constant [23]. For example, the rate constant for Lyn binding to the phosphorylated β ITAM ($k^*_{+L}$) is independent of whether Lyn or Syk is bound to any of the other sites within a receptor aggregate and is used to characterize 144 distinct chemical reactions. Since the rate of each reaction in the model is given by the product of the rate constant and the concentrations of the chemical species involved, the distribution of reaction rates within the same class mirrors the distribution of complexes that can participate in the reaction class. Just as the 164 Syk*-containing complexes represent alternative output channels of the model, the multiple reactions within each class represent alternative conduits of flow. The 17 reaction classes considered in our analysis and a breakdown of their rate distributions for the default parameter set are given in Table 1. The number of important reactions within each class (defined, as above, by a 95% threshold) is always a small fraction of the total number of reactions within a class. Cumulatively, only about 10% of the reactions in the network are characterized as important. A similarly narrow distribution of reaction rates is observed when the Lyn concentration is increased ten-fold (results not shown).

***Distribution of activation paths.*** Our final measure of network activity is the steady-state distribution of flux among reaction paths from inactive to activated Syk. Such a path is a non-repeating ordered sequence of reactions that transforms unphosphorylated cytosolic Syk into Syk*. The number of theoretically possible activation paths grows exponentially as a function of path length and far exceeds the number of molecules in the system (Table 2), but only 12 paths account for 50% of the total activation flux and ~1000 paths account for 95%. The top two paths (Fig. 3), both involve Syk binding to a receptor that is already bound to Syk. Such shortcutting paths minimize the opportunity for branching and are thus a major contributing factor to the narrow distribution of path flux. Path 54 (Fig. 3) has the highest flux among activation paths in which Syk initially binds to a complex containing no associated kinases, which cumulatively account for only 4% of the total activation flux. Thus, the full sequence of events illustrated in the path of Fig. 1C is not recapitulated for the vast majority of Syk activation events under steady state conditions. The species along the top two paths of Fig. 3 also exhibit the split levels of Lyn association that were observed in the top two Syk* complexes shown in Fig. 2. Increasing the Lyn concentration 10-fold dramatically reduces the flux of activation paths (values shown in parentheses in Fig. 3) involving complexes without Lyn (Paths 2 and 54).

***Variation of parameter values.*** To test whether a narrow distribution of network activity depends on parameterization of the model, we examine the three measures of the activity distribution for different sets of randomly altered parameter values. The level of Syk activation varies widely among parameter sets (Fig. 4A), but all parameter sets yield narrow distributions of network activity in comparison to a uniform distribution into all possible Syk*-containing species, reactions, or Syk activation paths (Fig. 4B-D). For 2-fold variations of parameters, each measure of activity is symmetrically distributed about the value characteristic of the original parameter set. For 10-fold variations of parameters, the average value of each measure decreases, although each distribution has a long tail that extends to higher values (Fig. 4B-D).

Systematic variation of parameter values confirms the example of Figure 2B: the identity and relative contribution of important network elements can change depending on parameter values (Fig. 4E-G). Figure 4E shows how the fractional contribution of Species 354, the species containing the highest concentration of Syk* using the original parameter set, is distributed in the "2x" and "10x" parameter set ensembles. Species 354 contains ~30% of the Syk* using the original parameter set (Fig. 2A-B), and its fractional contribution is distributed symmetrically about this value in the 2x ensemble over a range of ~10–60%. However, in the 10x ensemble, the distribution changes substantially, with the most frequent value of the fractional contribution tending towards zero (i.e., no Syk* in this state). The fractional contributions of the Syk autophosphorylation reaction with the highest reaction rate (Fig. 4F) and the Syk activation path with highest relative flux (Fig. 4G) exhibit similar behaviour. Thus, the relative contribution of an important network element is robust to modest (2-fold) parameter variations, but larger (10-fold) parameter variations usually cause activity to shift elsewhere in the network.

*Model reduction.* If a relatively small portion of the signalling network is active, one might expect that the FcεRI model could be reduced in size without changing its predictions. We tested this idea by removing species and their associated reactions to reduce the network size while minimizing the error of six specified output functions in comparison to the predictions of the full model. Permitting a maximum root-mean-squared (RMS) relative error of 10%, the smallest network we found contained 49 species and 118 reactions (Table 3). Although predictions of this model match those of the full model for the original parameter values, the reduced model is not predictive over a range of parameter values. Even for the 2x ensemble of altered parameter sets, the reduced model exhibits RMS errors outside the 10% tolerance in the vast majority of cases and exhibits >50% RMS error in a substantial fraction of cases. These results are insensitive to the size of the error tolerance used in model reduction (Table 3). The propensity of network activity to shift with parameter variations (Fig. 4) appears to limit the possibility of finding reduced models that apply over a broad range of cellular conditions.

## 4. Discussion

The protein-protein interactions of signal transduction [3], typified in the model considered here, generally imply a vast biochemical network, comprising a multitude of protein states and complexes and reactions among these. One issue that modellers of signal transduction must confront is whether this complexity affects the fundamental behaviour of the system or whether most of it may be safely ignored, as is common practice. The formulation of a simplified model amounts to assuming that a small number of states can effectively represent a multitude of potential states. One problem with such simplifying assumptions, aside from questions of accuracy, is that they limit the ability of models to predict the effect of typical experimental manipulations, such as knocking out specific sites of phosphorylation or domains of proteins.

We have attempted here to assess the role of molecular diversity in signal transduction by characterizing the diversity of complexes, reactions, and activation pathways that arise in a detailed model of early signalling events in a particular pathway initiated by receptor aggregation. We find that for any given state of the cell, characterized by a particular set of model parameters, only a small fraction of the network appears to be active (Fig. 2 and Table 1), but changing the cell's state can change which elements are active. The spectrum of active complexes in the model can be shifted dramatically even by a change in the concentration of a single protein (Fig. 2B and Fig. 3). Random variation of the model parameters demonstrates that the narrow distribution of network activity is a robust feature of the model (Fig. 4). The set of important network elements is generally robust to modest (2-fold) perturbations of rate constants and concentrations, and major shifts in activity require large (10-fold) variations. It is possible to find reduced models that reproduce the behaviour of the full model for particular parameter values, but, the predictions of these models are poor for perturbed cellular states (Table 3). They cannot be expected to predict accurately, for example, the effects of knocking out a particular protein domain. We conclude, therefore, that the assumptions of simplified models should be carefully validated before such extrapolations are made. The results of model reduction suggest that it will be difficult to find simplified models that are predictive over a broad range of cellular states.

This study also demonstrates that network dynamics alone, even in the absence of feedback or cooperative interactions, can produce highly focused flows of mass and information in a signalling network. Moreover, we have found that these flows can be regulated by parameters such as protein expression levels and enzymatic activities. One might expect such focused flows to arise from other mechanisms, such as cooperativity, feedback, or localization. These mechanisms may well restrict the range of complexes that form during response to a signal, but observation of limited molecular diversity among signalling complexes cannot be attributed to any particular mechanism without models that incorporate all of the potential mechanisms for limiting diversity. In particular, interpretation of proteomic data [9-13], assays of the protein phosphorylation states and complexes generated during signalling, will require models of the type analysed here to obtain mechanistic insights.

Experimental evidence for the role of differential complex formation in shaping cellular responses comes from studies of kinetic proofreading in immunoreceptor signalling (recently reviewed in [4]), which indicate that the signalling properties of a ligand are sensitive to the lifetime of ligand-receptor binding. Ligands with longer association lifetimes tend to signal more effectively because they generate "mature" signalling complexes that carry the signal downstream, whereas shorter binding ligands produce "frustrated" complexes that do not signal and can actually inhibit the production of mature complexes by sequestering signalling components in limited supply. Such "antagonist" ligands have been shown to produce both altered patterns of receptor phosphorylation [36] and kinase-sequestering complexes that inhibit signalling by more strongly binding ligands [37, 38]. Both of these effects are predicted by detailed models of early signalling events [35, 39], which provide theoretical support for the ideas incorporated in simplified models of kinetic proofreading [40, 41]. In terms of the potential role that differential complex formation may play in determining and regulating signalling outcomes, these effects represent just a few possibilities. Investigating these should be a major focus of computational studies of signal transduction [6, 42, 43] in the near future.

We have shown here that the pattern of complexes formed during a response to a signal can be sensitive to quantitative parameters that define the initial state of the cell. Because the spectrum of active complexes in our model can be shifted dramatically, even by a change in the concentration of a single protein, one function of the combinatorial complexity found in signalling systems might be to provide a mechanism for cellular decision making. Any event that changes the expression level or activity of a component of the cell could affect signal processing through a cascade involving that component by changing the composition of signalling complexes that are generated. In this way, the complexity of signalling complexes, which until now has been merely perplexing, might turn out to be an essential element of cellular computation.

## 5. Acknowledgments

We dedicate this paper to the memory of Carla Wofsy. Thanks to Dan Coombs, Tony Redondo, and Henry Metzger for helpful discussions. This work was supported by grants GM35556 and RR18754 from the National Institutes of Health and by the Department of Energy through contract W-7405-ENG-36.

**Figure captions**

**Fig. 1.** Model for early events in signal transduction through the FcεRI receptor. **(A)** The four basic components of the model—a bivalent ligand, the FcεRI receptor, and the kinases Lyn and Syk. Covalently cross-linked IgE dimers are bivalent ligands that bind and aggregate receptors irreversibly on the time scale considered in the model. The receptor is composed of three distinct subunits, the extracellular α subunit that binds the Fc portion of IgE with 1:1 stoichiometry, and the cytoplasmic β and $\gamma_2$ subunits that contain immunoreceptor tyrosine-based activation motifs (ITAMs), which upon phosphorylation bind to the SH2 domains of Lyn and Syk respectively. Lyn also associates with the unphosphorylated β subunit through an interaction involving its N-terminal unique domain. **(B)** Coarse description of the events leading to Syk activation in the model. **(C)** A sequence of reactions in the model that generate the receptor dimer complex with the highest stoichiometry of binding partners and phosphorylation. This path is one of the multitude of paths that exist in the model because of the large number of branches that exist at each step. **(D)** The distribution of the number of possible reactions that species containing a dimer of receptors can undergo. There are 300 such species in the model.

**Fig. 2.** Predicted distribution of activated Syk (Syk*) after introduction of IgE dimer (10 nM) at time $t=0$ s. Calculations were performed using the BioNetGen software package [29] using parameter estimates for the RBL-2H3 cell line [23], except as noted below. **(A)** Time courses for the total amount of Syk* (black curve) and the amount of Syk* in each of the two species containing the most Syk* at $t=100$ s (red and blue curves). **(B)** Rank ordered distribution of Syk* at $t=100$ s (black bars) and when the Lyn

concentration is increased 10-fold (red bars). The 12 complexes indicated account for more than 95% of the Syk* at $t=100$ s, and five of these account for 95% of the mass when the Lyn concentration is increased 10-fold. The indices used to refer to complexes are defined at our web site (http://cellsignaling.lanl.gov). **(C)** Illustration of the four species containing the most Syk* at $t=100$ s.

**Fig. 3.** Reaction paths that convert inactive cytosolic Syk to the activated form (Syk*) under steady-state conditions. The paths are indexed by the rank of their relative flux, which is given as a percentage of the total activation flux, the rate of Syk* turnover. The relative flux of each path when the total Lyn concentration is increased 10-fold is shown in parentheses. Relative fluxes are determined from a sample of $10^7$ randomly generated successful activation sequences.

**Fig. 4.** Effect of random variation of model parameter values on the distribution of network activity. The distributions of Syk* in the output species and of reaction rates grouped by rate constant are determined at $t=100$ s following stimulation with IgE dimer. The distribution of activation path fluxes is sampled at $t=1000$ s. For each of the following properties the panels plot relative frequency of occurrence in the 2x (solid lines) and 10x (dashed lines) ensembles: **(A)** Total level of Syk*. **(B)** Number of important Syk* species (account for more than 95% of Syk*). **(C)** Number of important reactions (carry more 95% of the reaction flux in all reaction classes, as defined in Table 1). **(D)** Number of important Syk activation paths (carry more than 50% of the Syk activation flux, as determined from a sample of $10^5$ activation events for each parameter set). **(E)** Fraction of Syk* contained in Species 354 (see Fig. 2C). **(F)** Fraction of Syk activation due to the Syk autophosphorylation reaction with the highest flux using the original parameter values. **(G)** Fraction of Syk activation due to Path 1 of Fig. 3. Filled circles on the *x*-axis indicate the value of each property calculated using the original parameter values.

**Figure 1**

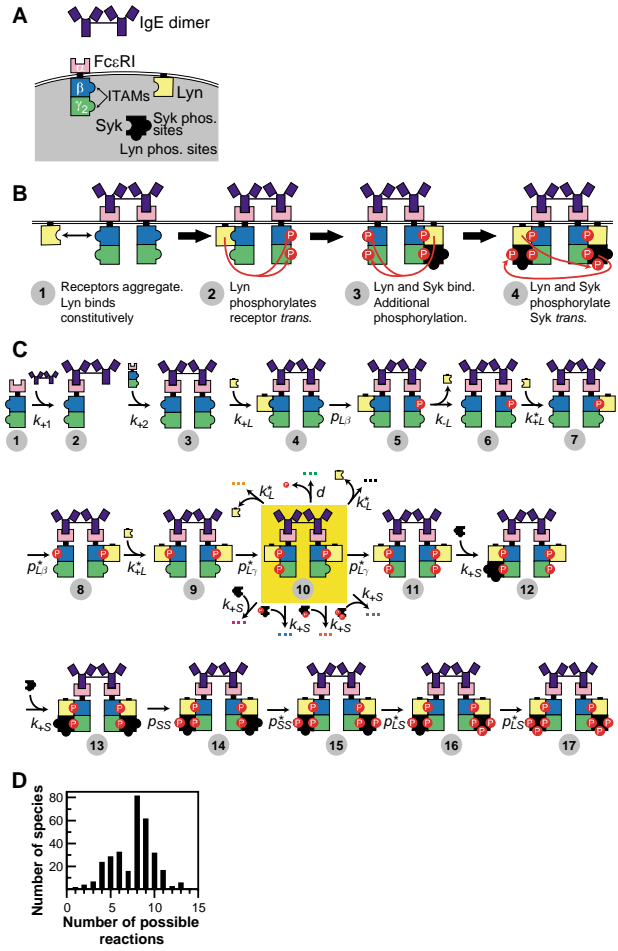

**Figure 2**

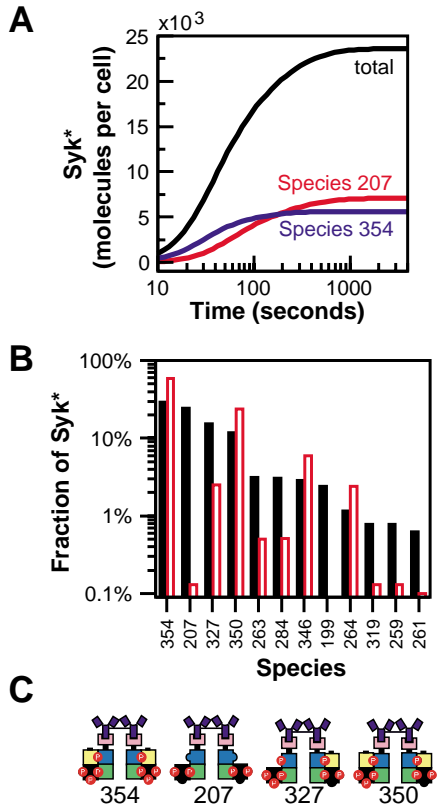

**Figure 3**

Path 1–9.2% (17.0%)
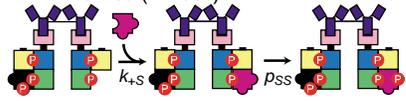

Path 2–7.7% (0.8%)
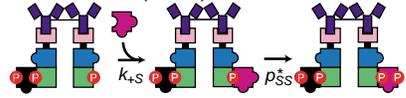

Path 54–0.2% (0.09%)
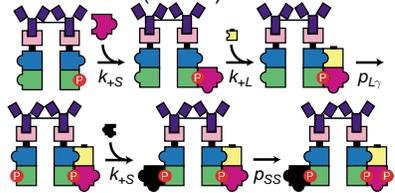

**Figure 4**

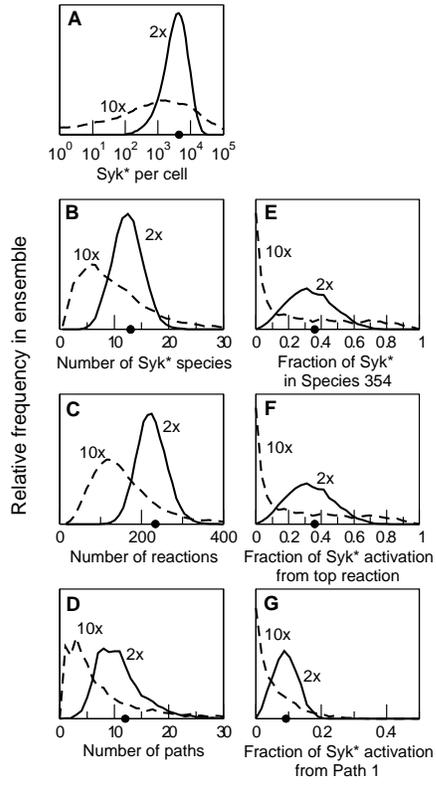

**Table 1.** Distribution of reaction rates for the RBL-2H3 parameter set 100 s after stimulation with 10 nM IgE dimer.

| Reaction Class[a] | Rate constant[a] | Number of reactions | Number of important reactions | Relative rate (% of total) All reactions in class | Relative rate (% of total) Top reaction in class |
|---|---|---|---|---|---|
| Ligand binding | $k_{+1}$ | 2 | 1 | 0.03 | 0.03 |
| Receptor aggregation | $k_{+2}$ | 4 | 1 | 0.03 | 0.03 |
| Constitutive Lyn binding | | | | | |
|   Association | $k_{+L}$ | 146 | 3 | 8.58 | 6.55 |
|   Dissociation | $k_{-L}$ | 146 | 6 | 8.59 | 6.55 |
| Lyn recruitment | | | | | |
|   Association | $k^*_{+L}$ | 144 | 19 | 0.11 | 0.03 |
|   Dissociation | $k^*_{-L}$ | 144 | 26 | 0.10 | 0.04 |
| Syk recruitment | | | | | |
|   Association | $k_{+S}$ | 384 | 20 | 0.27 | 0.06 |
|   Dissociation | $k_{-S}$ | 384 | 35 | 0.26 | 0.09 |
| Phosphorylation | | | | | |
|   Lyn→β ITAM | $p_{L\beta}$ | 36 | 5 | 1.70 | 1.14 |
|   Lyn*→β ITAM | $p^*_{L\beta}$ | 36 | 9 | 4.10 | 1.63 |
|   Lyn→γ ITAM | $p_{L\gamma}$ | 24 | 4 | 0.08 | 0.04 |
|   Lyn*→γ ITAM | $p^*_{L\gamma}$ | 24 | 7 | 0.18 | 0.08 |
|   Lyn→Syk | $p_{LS}$ | 48 | 8 | 0.35 | 0.14 |
|   Lyn*→Syk | $p^*_{LS}$ | 48 | 12 | 13.42 | 6.12 |
|   Syk→Syk | $p_{SS}$ | 64 | 11 | 2.02 | 0.63 |
|   Syk*→Syk | $p^*_{SS}$ | 64 | 10 | 19.17 | 6.10 |
| Dephosphorylation | $d$ | 628 | 53 | 41.00 | 6.07 |
| **Total** | 17 rate constants | 2326 | 230 | 100.00 | 35.33 |

[a] Complete definitions of reaction classes and rate constants are in Ref. [23].

**Table 2.** Number of possible paths and frequency of observed Syk activation paths as a function of path length. The number of observed paths and fraction of the total activation flux accounted for by paths of a given length are determined by stochastic sampling of $10^7$ successful activation events at steady state, when all FcεRI are aggregated into dimers.

| Path length | Number of possible paths | Number of observed paths | Fraction of activation flux |
|---:|---:|---:|---:|
| 2 | 64 | 64 | 38.2% |
| 3 | 384 | 287 | 26.3% |
| 4 | 2,056 | 773 | 12.6% |
| 5 | 14,068 | 1,434 | 4.8% |
| 6 | 108,728 | 1,831 | 4.4% |
| 7 | 845,800 | 2,026 | 4.3% |
| 8 | 6,301,796 | 2,204 | 3.3% |
| 9 | 44,621,932 | 3,081 | 2.1% |
| 10 | 300,913,268 | 4,206 | 1.3% |
| **Total** | 352,808,096 | 15,906 | 97.3% |

**Table 3.** Performance of reduced models measured by the RMS error of six observables (FcεRIβ ITAM phosphorylation, FcεRIγ ITAM phosphorylation, Syk linker region phosphorylation, Syk kinase activation loop phosphorylation, low-affinity Lyn-receptor association, and high-affinity Lyn-receptor association) at three time points ($t$=10, 100, and 1000 seconds). Results are representative of at least three reduced models with the same number of nodes produced by separate optimization runs.

**44 state model (104 reactions)**

|                        | Default set | 2x ensemble | 10x ensemble |
|------------------------|-------------|-------------|--------------|
| Mean RMS error         | 6.5%[a]     | 56%         | 50,000%      |
| % sets RMS error <10%  | –           | 4.6%        | 0.2%         |
| % sets RMS error >50%  | –           | 28%         | 77%          |

[a] Error with 1 nM IgE dimer stimulation. Error is 10.0% when objective function is evaluated at conditions under which model reduction was performed (10 nM IgE dimer stimulation, objective function computed at $t$=1,10, 100, and 1000 s).

**83 state model (257 reactions)**

|                        | Default set | 2x ensemble | 10x ensemble |
|------------------------|-------------|-------------|--------------|
| Mean RMS error         | 3.0%[a]     | 45%         | 120%         |
| % sets RMS error <10%  | –           | 16%         | 1.8%         |
| % sets RMS error >50%  | –           | 22%         | 54%          |

[a] Error with 1 nM IgE dimer stimulation. Error is 1.0% when objective function is evaluated at conditions under which model reduction was performed (10 nM IgE dimer stimulation, objective function computed at $t$=1,10, 100, and 1000 s).

## Appendix 1: Enumeration and sampling of activation paths

We define an activation path as a sequence of reaction events by which a molecular component of the model is transformed from an inactive state into an active one. Here, we focus on paths that transform an unphosphorylated Syk molecule in the cytosol into in an autophosphorylated Syk molecule associated with a receptor dimer complex (Syk*), but the methods can be easily generalized. The full reaction network is first transformed into a directed graph (a set of nodes and directional edges connecting nodes), from which activation paths are defined, enumerated, and sampled to determine relative activation fluxes.

*Constructing the component activation graph.* Each node in this activation graph represents a distinct state of Syk in the model. Nodes are created from the species that contain Syk; species that contain one Syk molecule give rise to one node, but species that contain multiple Syk molecules in distinct states give rise to multiple nodes. For example, in the second species of Path 1 in Figure 3, the labelled Syk may be associated with either the left or the right receptor of the complex. Thus, to account for both possibilities, we must include two nodes in the graph for this species. The edges of the activation graph correspond to directed chemical transitions between nodes that can be carried out in a single reaction step. Edges are created from the reactions in the model that involve Syk; one edge is created for each distinct pair of reactant and product nodes arising from the reaction. Reactions that contain multiple Syk molecules give rise to multiple edges. For example, the first reaction shown in Path 1 of Figure 3, where the labelled Syk may be either cytosolic (the purple Syk) or associated with a receptor complex (the black Syk), gives rise to two edges. The weight of an edge is given by the rate at which a molecule of the labelled Syk is transformed by the reaction. (If multiple reactions carry out the same transformation of nodes, the weight is the sum of the relevant rates.) Fro the example given above, the weight of the edge involving transformation of the purple Syk is given by $k_{+S}$ times the concentration of the species containing the black Syk, whereas the weight of the edge involving the transformation of the black Syk is given by $k_{+S}$ times the concentration of free Syk in the cytosol. (Note that these weights are in general time dependent.) Reactions involving the loss of Syk from a symmetric complex can give rise to two edges from a single reactant node. If the number of Syk molecules in the complex is $s$, the weight of the edge for dissociation of the labelled Syk from the complex is $1/s$, and the weight of the edge for retention of the labelled Syk is $(s-1)/s$. The Syk activation graph constructed in this manner contains 420 nodes, of which 192 represent activated states of the labelled Syk, and 3644 non-zero edges (for irreversible ligand binding).

*Formal definition of an activation path.* A Syk activation path is defined as an ordered sequence of nodes of this graph, where the first node corresponds to unphosphorylated, cytosolic Syk, and the final node is the first node in the sequence in which the labelled Syk is autophosphorylated and part of a receptor dimer complex. Each pair of adjacent nodes in this sequence must be connected by an edge with non-zero weight. To simplify our analysis, we restrict the definition of a path to include only those sequences in which each node appears at most one time to avoid cycles within paths.

*Enumeration of paths.* The enumeration of possible paths as a function of path length (column 2 of Table 2) is carried out using a modified form of the depth-first search [44]. Paths up to length *N* are enumerated as follows. A path is implemented as a stack (elements are added to and removed from the end of the list) and is initialized with a starting node corresponding to unphosphorylated, cytosolic Syk. (*I*) loop over the edges originating from the final node of the path. If the final node of an edge corresponds to an active state of the labelled Syk, increment the number of paths of length *n*, where *n* is the number of nodes in the path, and continue with loop (*I*). If *n*<*N* and the final node of the edge is not a member of the current path, add this node to the path and begin a new loop at (*I*). (*II*) When the edges from the final node in a path are exhausted, remove this node from the path and continue with loop (*I*) if the path still contains at least one element. The recursive looping implied by (*I*) is implemented using a second stack that contains a pointer to the current edge for each node in the path stack, where the edges for each node are stored in a linked list and looped over in that order.

*Determination of activation flux.* Sampling of paths to determine their relative contribution to the total activation flux is done using a stochastic algorithm based on Gillespie's method for computing chemical dynamics [45]. The first node in a path is unphosphorylated, cytosolic Syk. Paths are extended from the terminal node *i* in the path sequence by choosing the next node *j* randomly with probability $p_{i \to j} = w_{i \to j} / \sum_{j'} w_{i \to j'}$, where $w_{i \to j}$ is the weight of the edge taking *i* into *j*. Paths are terminated when the Syk molecule being traced is autophosphorylated (successful activation) or when it returns to the cytosol in its unmodified state. Sampling of paths continues until a specified number of activation events is recorded. Following a successful trace, the path is pruned to remove loops by iteratively removing all nodes between repeating nodes (including one instance of the repeating node) until no more repeating nodes are present in the path. The relative activation flux from a given activation path $p$ is $\frac{\text{\# times } p \text{ observed}}{\text{\# activation events}}$. Edge weights are determined from the species concentrations at a particular sampling time and assumed to be constant. This assumption is equivalent to assuming that the relative change in the species concentrations is small over the duration of the activation events being sampled, which holds exactly under the steady state conditions used to generate the data in Figure 3 and Table 2. This assumption is also reasonably accurate for the $t=1000$ s time point used in the sampling of activation paths in the parameter set ensembles, because the duration of the vast majority of activation events is at most a few seconds and species concentrations undergo small fractional changes on this time scale at a time so distant from the addition of ligand.

## Appendix 2: Algorithm for model reduction

The optimization algorithm is as follows. The starting network is taken to be the full network. (1) At each step a move is attempted in which a species is randomly deleted from the network. If the objective function (defined in Methods) is below the threshold value for the RMS, the deletion is accepted and the optimization continues from (1). (2) Following a failed move, the deletion is removed from the list of possible deletions from the current network. (a) After a sequence of 50 failed deletions or when all deletions from the current network have been exhausted, 2 addition moves are made in which a randomly selected species that was previously deleted is added back to the network. Additions are allowed only if they do not increase the value of the objective function. (b) After a sequence of 1000 moves in which the size of the smallest network reached has not decreased, 50 addition moves of type (a) are performed. Optimization then continues from (1). The purpose of both types of addition moves, which undo the effects of past moves, is to prevent the procedure from being trapped in a local minimum.

We varied both parameters and procedures of this optimization algorithm, but found that the above recipe produced the smallest reduced networks for a given value of the objective function threshold, with the smallest spread in the size of the smallest network found from different optimization runs. For example, networks with 44 nodes that satisfied an error tolerance of 10% were found in 3 of 16 optimization runs consisting of about $10^6$ attempted moves each. The range in the size of the smallest network found in these 16 runs was 44-49. Similarly, 4 of 16 optimization runs with an error tolerance of 1% found reduced networks with 83 nodes, and the range in the size of the smallest network found was 83-90. Reduced models are available from the authors upon request.